\documentclass[useAMS,usenatbib]{mn2e}

\def\aap{AA}
\def\apjl{ApJL}
\def\apjs{ApJS}
\def\mnras{MNRAS}
\def\apj{ApJ}
\def\aj{AJ}

\def\pasj{PASJ}
\def\nat{Nat}

\usepackage{graphicx}
\usepackage{float}
\usepackage{bm} 
\usepackage{amssymb}
\usepackage{amsmath} 
\usepackage[export]{adjustbox}
\usepackage{mathrsfs} 
\usepackage{mathtools}
\usepackage{color}

\title[An atlas of stellar deposition]{Contributions to the accreted stellar halo: an atlas of stellar deposition}
\author[N. C. Amorisco]{N. C. Amorisco$^{1,2,3}$\thanks{E-mail: nicola.amorisco@cfa.harvard.edu} \\
$^{1}$Max Planck Institute for Astrophysics, Karl-Schwarzschild-Strasse 1, 85748 Garching, Germany\\
$^{2}$Institute for Theory and Computation, Harvard University, 60 Garden Street, Cambridge, MA 02138, USA\\
$^{3}$Dark Cosmology Centre, Niels Bohr Institute, University of Copenhagen, Juliane Maries Vej 30, DK-2100, Copenhagen}

\begin{document}



\maketitle

\label{firstpage}

\begin{abstract}
The accreted component of stellar halos is composed of the contributions of several satellites,
falling onto their host with their different masses, at different times, on different orbits. This work uses a suite of 
idealised, collisionless N-body simulations of minor mergers {and a particle tagging technique}
to understand how these different ingredients shape each contribution to the accreted halo, in both density and kinematics.
I find that more massive satellites deposit their {stars} deeper into the gravitational potential
of the host, with a clear segregation enforced by dynamical friction.
Earlier accretion events contribute more to the inner regions of the halo; more concentrated subhaloes sink deeper 
through increased dynamical friction. 
The orbital circularity of the progenitor at infall is only important for low-mass satellites: dynamical friction efficiently radialises 
the most massive minor mergers erasing the imprint of the infall orbit for satellite-to-host virial mass ratios $\gtrsim1/20$. 
The kinematics of the stars contributed by each satellite is also ordered with satellite mass: low-mass satellites contribute fast-moving  
populations, in both ordered rotation and radial velocity dispersion. In turn, contributions 
by massive satellites have lower velocity dispersion and lose their angular momentum to dynamical friction, 
resulting in a strong radial anisotropy.

\end{abstract}


\section{Introduction}

The abundance of substructure that encircles the Milky Way \citep[e.g.][]{RI95, VB06, CG09},
Andromeda \citep[e.g.][]{RI07,AM09,JV14}, and nearby galaxies \citep[e.g.][]{MD08,MD10,At13,DP15,DC16} testifies 
that at least part of the tenuous stellar haloes that surround galaxies have been contributed hierarchically 
by smaller merging systems \citep[e.g.,][]{Egg62,SZ78,SW91,KJ08}. As a consequence, the detailed properties of 
individual stellar haloes are the result of a highly stochastic process. This is especially true for Milky Way 
like galaxies and for galaxies with lower virial mass, where the total budget of the accreted stellar halo
is predicted to be dominated by just a handful of satellites \citep[e.g.][]{BJ05,MA06,LS07}. 
For example, evidence is building up that the stellar halo of the Milky Way is quite different from the one of Andromeda, 
with the first featuring a clearly broken power-law density profile \citep[][although see also the recent Cohen et al. 2015]{AD11,AD14},
while the second displaying an apparently smooth and comparatively shallow density profile out to almost the virial radius \citep[e.g.,][]{KG12,RI14}.

The origin of this stochasticity lies in the fact that, even after full tidal disruption and complete mixing in phase space, 
both density and kinematic profiles of the accreted 
halo keep bearing the signature of the details of the assembly history of the host \citep[e.g.,][]{BJ05,AC10,AD13,AP14}. 
Just to mention the most important, these details include the masses of the main contributors and their internal structures, 
their infall times and infall orbits. All these ingredients play a role in determining where and how stars are deposited onto the host, 
in both density and kinematics, contributing to the stochasticity of the outer stellar halo in potentially different ways. 
However, a clear theoretical understanding of the role and relevance of these different physical ingredients is still largely missing.

Some correlations have emerged from both observational and theoretical studies. Using stacks of SDSS data, 
\citet{RDS14} observes that the density slope of the outer stellar envelope is correlated with the stellar mass of the galaxy, but
also that galaxies of different morphological types follow different average behaviours. \citet{AD13} uses the suite of
simulations produced by \citet{BJ05} to study what kind of accretion histories result in broken density profiles. Though on a rather small
sample, they find hints that breaks are more evident in case of quiet recent accretion histories.
Finally, \citet{AP14} perform a thorough study of the outer haloes in the Illustris simulations \citep{MV14}, 
and find that average density slopes show a significant correlation with both the total virial mass of the host
and its formation time: less massive hosts that assemble earlier on display steeper stellar haloes.

Without aiming to produce realistic or detailed models of the stellar halo, I concentrate here on systematically exploring  
the role of satellite mass, internal structure, infall redshift and infall orbit on shaping the individual contributions to the accreted stellar
halo. The objective of the present paper is to investigate where, in terms of orbital energy within the host, different satellites deposit 
the bulk of their stars. Counterbalancing mechanisms at work are tidal stripping and dynamical friction. Satellites infall on energetically 
similar, loosely bound orbits \citep[e.g.,][]{AB05,AW11,LJ15}, { and then have their energy and angular momentum consumed by 
dynamical friction, sinking gradually towards the center of the host. At each time, the drag force is instantaneously stronger for more massive 
remnants, although it can operate for longer times on the less massive subhaloes, which are more resilient to stripping 
because of their higher concentration. }
This makes the question of where each satellite releases and deposits its stars while sinking towards the depths of the host potential 
a non trivial one. 

{Both tidal stripping and dynamical friction have been widely studied in the literature, as they're vital
to a wide range of astrophysical problems. The post-infall evolution of satellite galaxies  
has been the subject of a number of recent works, both with reference to their subsequent morphological transformation \citep[e.g.,][]{LM01a,SK11,To16} 
to the tidal stripping of their stellar/dark constituents \citep[e.g.,][]{AK04,JP08,Ch13,DE15},
and to the surviving subhalo populations of hosts within a fully cosmological framework
\citep[e.g.,][]{LG04,JD04,MK07,VS08,AL09}}. {Recently}, \citet{FB15} have provided 
a comprehensive analysis of the significance and extent of segregation in the properties of surviving subhalos,
both in terms of their spatial distribution and of their orbital energy. Halo substructure is strongly segregated
as a function of the accretion redshift, as the more recently accreted haloes have not yet had the time to
sink in. As a direct consequence, subhaloes that are closer to centre of the host (or are on more bound orbits)
have lost a higher percentage of their initial mass \citep[e.g.][and references therein]{FB15}.

By definition, the works just mentioned concentrate on the properties of the bound remnants, 
while, motivated by studying the connection with the properties of stellar haloes, here I focus on the tidally shed 
material, and on whether their kinematics within the host is clearly correlated
with the properties of that specific accretion event. In order to address this, I use idealised N-body simulations 
of minor merger events. A suite of purely collisionless runs is constructed so as to cover those regions of the 
parameter space that are representative of accretion events in a $\Lambda$CDM universe, and that collect the
major contributions to the accreted stellar halo.

From the point of view of the methodology, \citet{BK08} have also performed a suite of idealised simulations of 
minor mergers with different mass ratios and orbital properties. However, their attention was focussed on measuring the time 
scale of these mergers, and on its dependence on the initial mass ratio and orbital properties.

The structure of this manuscript is as follows: Section~2 presents the numerical setup and illustrates  
the suite of runs; Section~3 collects the results of this study in dimensionless units; Section~4 scales them to a Milky Way like galaxy, and 
contextualises findings; Section~5 concentrates on the kinematics of the deposited stars;
Section~6 provides a summary of the main findings and lays out the conclusions.

\section{Simulations}

\begin{table}
 \centering
 \begin{minipage}{80mm}
  \caption{Structural and orbital parameters of the suite of minor merger N-body simulations.}
  \begin{tabular}{@{}ccccc@{}}
  \hline
 &$M_{\rm vir, s}/M_{\rm vir, h}$  & $\log{{r_{0,{\rm s}}/r_{0,{\rm h}}}\over{(r_{0,{\rm s}}/r_{0,{\rm h}})_{\Lambda {\rm CDM}}}}$ & $r_{\rm circ}/r_{0,{\rm h}}$ & $j$ \\
\hline
A & 1/122 & 0  & 5 &\{0.2,0.5,0.8\}\\ 
B &1/50.0  & 0 & 5 & \{0.2,0.5,0.8\}\\ 
C& 1/20.4  & 0 & 5  & \{0.2,0.5,0.8\}\\ 
D&1/13.0  & 0 & 5 & \{0.2,0.5,0.8\}\\ 
E&1/8.33  & 0 & 5 & \{0.2,0.5,0.8\}\\ 
F&1/3.40  & 0 & 5 & \{0.2,0.5,0.8\}\\ 
\hline
G&1/50.0  &  $0\pm\sigma\times$\{1,2.5\}& 5 & 0.5\\ 
H&1/8.33  &  $0\pm\sigma\times$\{1,2.5\}& 5 & 0.5\\ 
\hline
I&1/50.0  & 0 & 9 & 0.5 \\
J&1/8.33  & 0 & 9 & 0.5  \\
\hline
\end{tabular}
\end{minipage}
\end{table}

As a working hypothesis, I assume that the total density distributions of both hosts and satellites are well described by  
spherically symmetric, {non-rotating} NFW density profiles \citep[][]{NFW97}.
Initial conditions for all simulations are generated from the phase space distribution function of an isotropic NFW structure, exponentially 
truncated at the virial radius, calculated using Eddington's inversion \citep{Edd16}, as delineated by \citet{Wi00}. As shown by \citet{SK04},
this procedure ensures long-term equilibrium, which the assumption of a locally Maxwellian distribution cannot guarantee.
All runs are collisionless N-body only simulations, executed using the publicly available code Gadget-2 \citep{VS05}. In all cases the satellite 
is populated with $N_{\rm s}=5\times 10^4$ particles, and all particles in the simulation have similar masses, so that the host is sampled 
with $N_{\rm h}\approx N_{\rm s}\ M_{\rm vir, h}/M_{\rm vir, s}$ particles. In the following, I will use the suffixes s and h to indicate, respectively,
quantities referring to the satellite and host.

{
\subsection{Stars and dark matter: particle tagging technique}

The simulations used here do not explicitly include stellar particles as a separate dynamical component of the merging satellite. 
Rather, I adopt a particle tagging technique, in which stars are represented by a fraction of the satellite's dark matter particles.
These are selected based on their binding energy within the satellite itself. Such particle-tagging strategy has proven to be a successful technique 
to study the properties of stellar haloes using dark-matter only cosmological simulations \citep[e.g., ][]{JB01,BJ05,AC10,AC13}. 
In particular, \citet{AC10} have shown that a viable tagging criterion should mainly select particles that are deeply embedded within the satellite's halo, 
as stars form close to the bottom of the gravitational potential. They show that the mass-size relation of galaxies is reasonably fit when stars
are identified with the most bound halo particles, up to a threshold $f_{\rm mb}$ of a few percent. Throughout this paper I adopt the 
nominal value of $f_{\rm mb}=5\%$. Appendix A explores the effect of reducing such threshold.

A particle-tagging strategy neglects all possible effects that baryonic physics could have on the dark haloes and on the stripping process. 
Here, for example, it is equivalent to assuming that minor mergers are dominated by a purely collisionless dynamics. The following Section 
builds on this assumption. I will discuss the limits of the tacit hypotheses that lie behind the particle tagging technique in some more detail in Sect.~6.1.}

\begin{figure}
\centering
\includegraphics[width=.9\columnwidth]{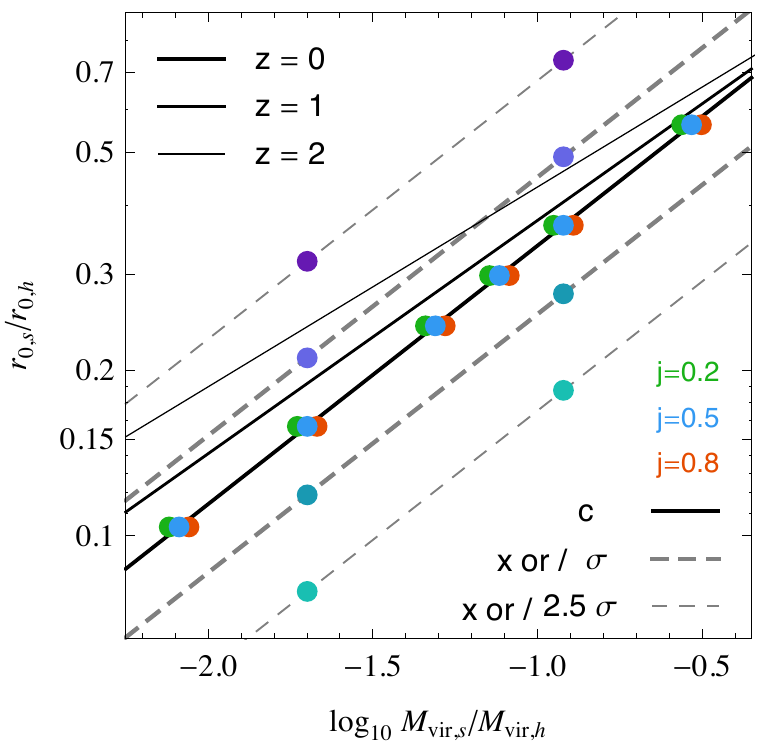}
\caption{The parameter space describing the structural properties of minor mergers between two NFW haloes:
mass ratio and ratio between the scale radii of the two halos, $r_{0,{\rm s}}/r_{0,{\rm h}}$. Full lines identify the median scaling 
relation between these two structural parameters in a $\Lambda$CDM universe at different redshifts, as given by the mass-concentration relation. Dashed lines show the effect of the 
scatter in the mass-concentration relation observed in cosmological simulations. Colored points illustrate how the suite of simulations performed here
samples this space (multiple points with different colours indicate runs with different initial orbital circularities). }
\label{parspace}
\end{figure}

\subsection{The parameter space}

In the approximation in which gravitation is the only force at play, in which the gravitational potential 
of both satellite and host is dark matter dominated, minor mergers  
can be described within a four-dimensional parameter space. Two dimensionless parameters determine the coupling 
between the structural properties of satellite and host, and two parameters set the orbital properties of the pair.

A convenient choice for the two structural parameters is the combination
\begin{equation}
\mathbf{\theta}_{\rm str}=\left({M_{\rm vir, s}\over M_{\rm vir, h}},\ { r_{0,{\rm s}}\over r_{0,{\rm h}}}\right)\ ,
\label{strpair}
\end{equation}
featuring the virial mass ratio between satellite and host, and the ratio between their characteristic
radii (equivalently, the latter structural parameter can be replaced by the ratio between the
characteristic densities of the two haloes, $\rho_{0,{\rm s}}/\rho_{0,{\rm h}}$).
Cosmological simulations have shown that the structural properties of $\Lambda$CDM haloes are highly correlated, so that
halo mass, concentration and redshift are closely related \citep[e.g.,][and references therein]{LG08,MC11,AL14}. 
This implies that minor merger events that are relevant to a $\Lambda$CDM cosmology do not populate the entire plane
defined by the parameters~(1), but only a well defined region of it.
In particular, as the mass-concentration relation at a fixed redshift is essentially scale-free (i.e. a power-law), the locus
that identifies relevant minor mergers is independent of the host halo mass $M_{\rm vir, h}$.

Full lines in Figure~1 show how the ratio $r_{0,{\rm s}}/r_{0,{\rm h}}$ scales with the virial
mass ratios $M_{\rm vir, s}/M_{\rm vir, h}$. The Figure concentrates on the range of mass ratios that the 
most important contributors to the stellar halo are expected to inhabit.
Lines of different thickness display the redshift evolution of this locus, for $z\in\{0,1,2\}$, i.e. the range of 
times in which most of the ex-situ stellar mass is expected to be accreted. At higher redshifts the concentrations of haloes of 
different mass are more similar to each other, implying that the structural ratio $r_{0,{\rm s}}/r_{0,{\rm h}}$ is closer to unity 
than what seen at the present epoch.
Dashed lines identify a measure of the characteristic scatter observed in cosmological simulations \citep[e.g.,][]{AL14}, 
by showing the effect of varying the concentration $c$ of the satellite by a factor of approximately 1 and 2.5 sigma (with respect to 
its average value at redshift $z=0$). At constant mass ratio, satellites that are less concentrated than average (or that infall 
on more concentrated hosts) result in a larger $r_{0,{\rm s}}/r_{0,{\rm h}}$, and viceversa.

The two parameters describing the initial orbit of the pair are, respectively, a measure of its energy and angular momentum. 
\citet{LJ15} show that the orbital energy of satellites at infall is comparable with the energy of the circular orbit with 
radius equal to the virial radius of the host: 
\begin{equation}
E_{\rm inf}= E_{\rm circ}(r_{\rm circ})\  \  {\rm with}\  \  r_{\rm circ}\approx c_{\rm h}r_{0,{\rm h}}\ ,
\label{orben}
\end{equation}
where $E_{\rm inf}$ is the orbital energy of the satellite at infall, $E_{\rm circ}(r_{\rm circ})$ is the energy of the circular orbit 
with radius $r_{\rm circ}$, and $c_{\rm h}$ is the host's concentration. A convenient measure of the initial orbital angular momentum 
is of course its circularity 

\begin{equation}
j\equiv J/J_{\rm circ}(E) \ ,
\label{circu}
\end{equation}
so that I adopt the pair of dimensionless parameters
\begin{equation}
\mathbf{\theta}_{\rm orb}=\left( {r_{\rm circ}(E_{\rm inf})\over r_{0,{\rm h}}},\ j\right)\ .
\label{orbpair}
\end{equation}
%

\subsection{The suite of runs}

Table~1 and Fig.~1 collect the details of the suite of runs. The main contributors to the stellar halo of Milky Way like galaxies
are expected to have a mass ratio of $M_{\rm vir, s}/M_{\rm vir, h}\sim 1/10$ \citep[e.g.][]{BJ05}, so that I explore the range 
$1/100\lesssim M_{\rm vir, s}/M_{\rm vir, h}\lesssim 1/3$. 
\begin{itemize}
\item{The main set of simulations (runs A to F) adopts the structural
ratio $r_{0,{\rm s}}/r_{0,{\rm h}}$ identified by cosmological simulations for the corresponding mass ratio at redshift $z=0$ (see Fig.~1). }
\item{A parallel set of runs (G and H) explores the
effect of the scatter expected in the structural properties of haloes. Note that such scatter fully includes any shifts 
due to the systematic change in the structure of haloes at different infall redshifts, at least while $z_{\rm inf}\lesssim2.5$. }
\item{I consider two values of the initial orbital energy $E_{\rm inf}$: $r_{\rm circ}/r_{0,{\rm h}}\in\{5,9\}$. These approximately contain 
the interval that is representative of accretions onto a Milky Way sized haloes at intermediate redshifts. Runs I and J are used to 
estimate the effect of this additional degree of freedom.}
\end{itemize}

Finally, cosmological accretions have circularities at infall that are centred on intermediate 
values, $j\approx 0.5$, but with a significant spread towards both radial and circular orbits \citep[e.g., ][]{LJ15}. 
In order to explore the full allowed range, I consider the cases $j\in\{0.2,0.5,0.8\}$ (runs A to F). As orbital energy stays fixed, each simulation
begins with the satellite at different apocentric distances.
In turn, before the effect of dynamical friction, these orbits have very similar orbital times, which are mainly a function of energy. 

All simulations are run for a total time that corresponds to 15 Gyr when $M_{\rm vir, h}$ is scaled to $10^{12}M_{\odot}$.
It is worth mentioning that not all satellites are entirely destroyed by this time. The most massive satellites I consider here are quickly disrupted within a 
couple of pericentric passages, but the lowest mass ones survive for much longer and are considerably more resilient
to tides as a result of the much higher contrast between their central density and the density of the host. 
For example, for all circularities, even after 15 Gyr, the satellites in runs A, with the lowest mass ratio, 
still display a bound nugget of about a few percent of the initial virial mass. This is also true for the highest-circularity 
and highest-concentration cases of run B. At the resolution used here, all other minor mergers are complete within 
the time interval covered by the studied runs. 

\begin{figure}
\centering
\includegraphics[width=.95\columnwidth]{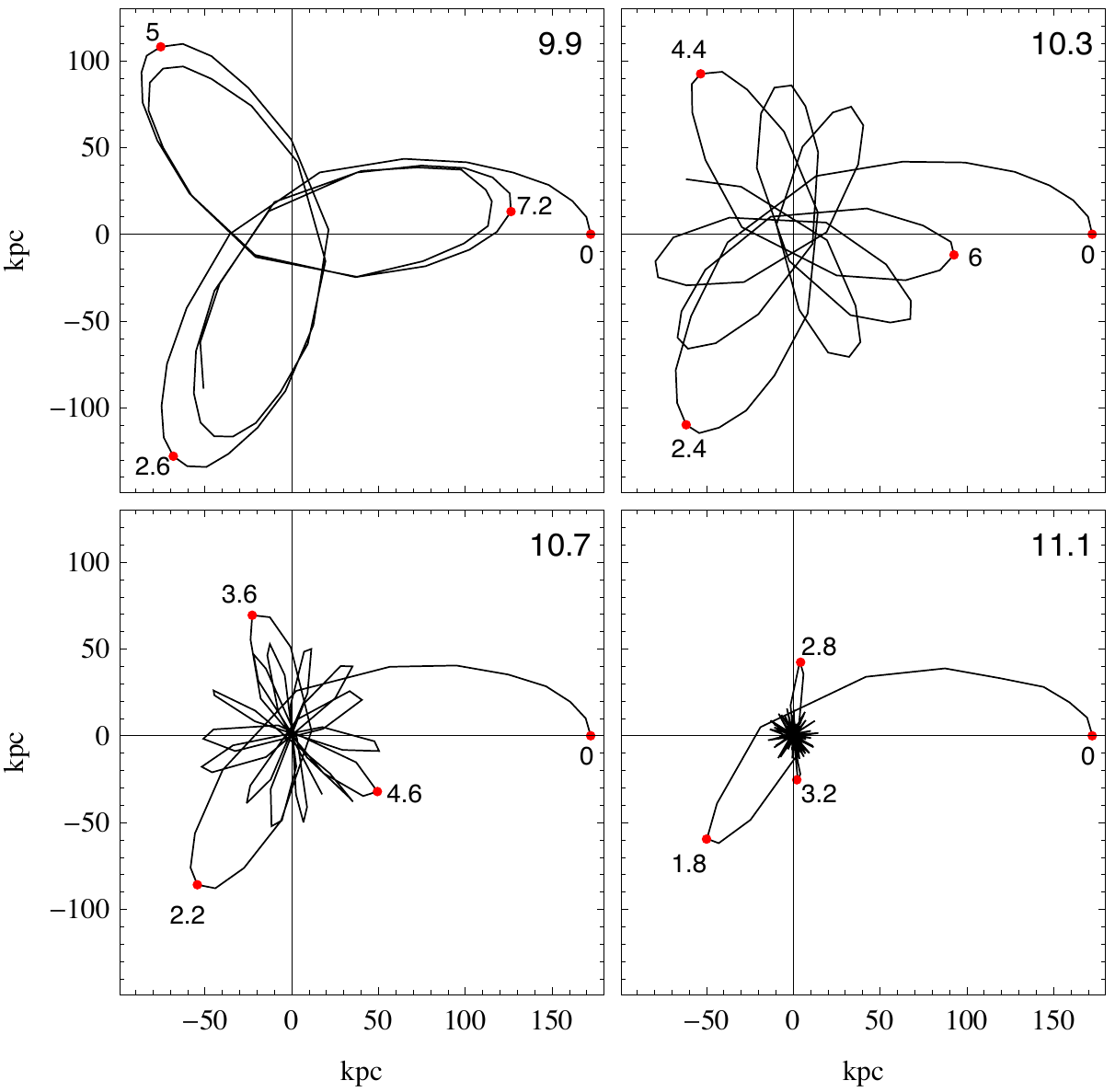}
\caption{The relative orbit of the satellite's and host's most bound particles. Satellite masses (indicated in the upper-right corner of each panel) have been
scaled to the case of a Milky Way like host, $M_{\rm vir, h}=10^{12}M_{\odot}$. The same physical scaling has been applied to the orbital
times, displayed for the sequence of the first four apocenters. All panels refer to the case $j=0.5$, with ${{r_{0,{\rm s}}/r_{0,{\rm h}}}={(r_{0,{\rm s}}/r_{0,{\rm h}})_{\Lambda {\rm CDM}}}}$, runs A, B, C, E.}
\label{circularity}
\end{figure}

\section{Mass deposition}

A visual impression into the unfolding of the minor mergers is displayed in Figure~2,
which shows the relative orbital distance between the satellite's and host's most bound particles,
for the case $j=0.5$ and ${{r_{0,{\rm s}}/r_{0,{\rm h}}}={(r_{0,{\rm s}}/r_{0,{\rm h}})_{\Lambda {\rm CDM}}}}$ (runs A, B, C and E).
To simplify the interpretation, displayed quantities have been scaled to the case of a Milky Way like host at redshift $z=0$, with $M_{\rm vir, h}=10^{12}M_{\odot}$
and $r_{0,{\rm h}}=21.1$kpc.
Each panel reports the logarithm of the satellite masses in the upper-right and the orbital times of the first four apocentric passages in Gyr.
The evolution in the apocentric distances shows that, in all cases, dynamical friction drags the satellite towards a more bound orbit. 
Energy is lost during some initial period of time, during which a massive-enough bound remnant is present, but the length of which 
depends on the initial satellite mass. Thereafter, energy is conserved until the end of the simulation, with no further orbital evolution. 

Fig.~2 is already showing that, all the rest being equal, 
dynamical friction is more effective on the more massive satellites. Although low-mass subhaloes survive for much longer,
their sink rate is too low to allow them to reach the central regions of the host. In fact, they are confined to the outskirts, where they
dissolve very slowly under the influence of tides.

\begin{figure*}
\centering
\includegraphics[width=\textwidth]{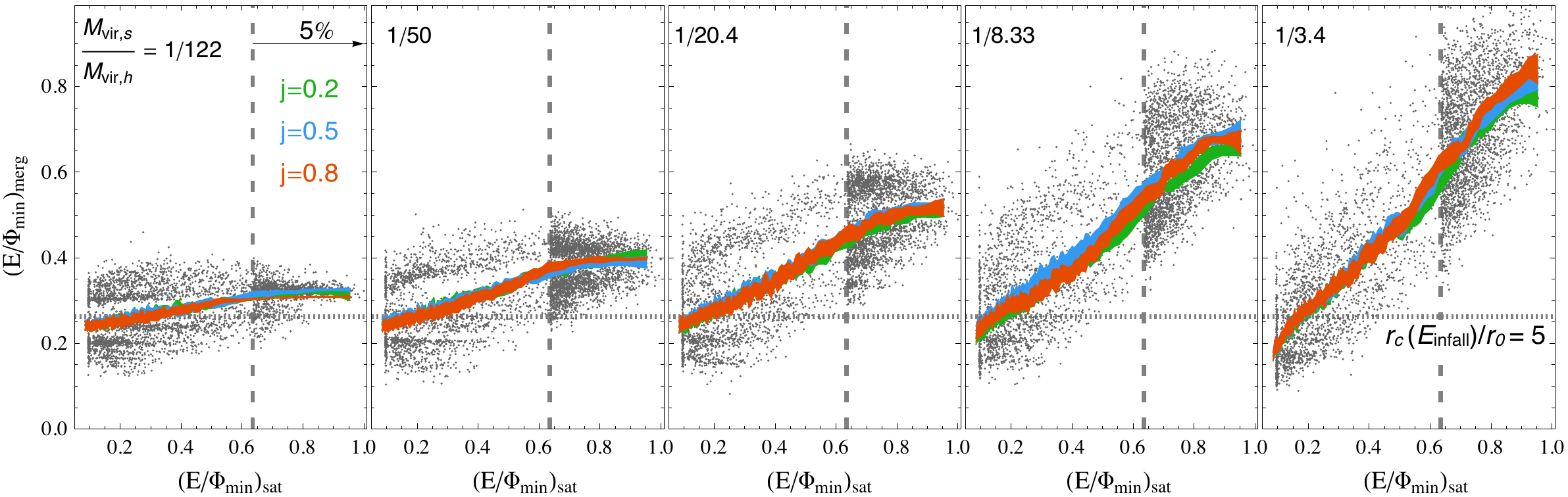}
\caption{The energy distribution of the satellites particles. The panels show the scatter plots of the final (normalised) energy of the satellite's particles within the host (on the $y$ axis), as a function of their initial (normalised) energy within the infalling satellite before interaction (on the $x$ axis). Higher values in both coordinates represent particles that sit deeper within the final host's or initial satellite's potential. Tidal stripping of the satellite proceeds in time from left to right, removing increasingly more bound particles. Dynamical friction operates dragging the satellite's remnant towards the top of each panel, deeper into the host. Panels illustrate the effect of different initial virial mass ratios, displayed in the top-left of each panel. Different colours in all panels are used to highlight median values (with one-sigma uncertainty) for infall orbits with different initial circularity: $j\in\{0.2, 0.5, 0.8\}$, as in the legend in the leftmost panel. Grey points always refer to the case $j=0.5$; for convenience, only 10\% of the dark matter particles are actually displayed (to the left of the vertical dashed  line identifying the stellar particles). Dynamical friction causes more massive satellites to deposit their {stars} deeper into the host's potential, with a clear gradient.}
\label{energy}
\end{figure*}

\subsection{Dynamical friction at play}

Figure~3 provides a quantitative view on the effect of dynamical friction. For different mass ratios, panels display scatter plots of the final 
energy of each satellite 
particle within the merger product against its initial energy within the satellite itself. More explicitly, $(E/\Phi_{\rm min})_{\rm merg}$ 
is the orbital energy of particles at the end of the simulation, normalised by the 
depth of the potential well. For example, the horizontal dashed line present in all panels of Fig.~3 identifies the orbital energy of the satellites
at infall. Particles with large values of $(E/\Phi_{\rm min})_{\rm merg}$ are more deeply bound within the merger 
product, up to $(E/\Phi_{\rm min})_{\rm merg}=1$, which identifies a particle sitting at rest at the centre of such a potential well. Analogously,
$(E/\Phi_{\rm min})_{\rm sat}$ is the normalised energy within the satellite before infall, and therefore measures how bound 
particles were within it before the merger event. {Stars sit at the bottom of the potential wells of the satellites, and are therefore to the
right end of all panels. In fact, the vertical dashed lines separate the 5\% most bound particles in the satellite itself. As 
described in Sect.~2.1, these are tagged as stars.} Tidal stripping proceeds from left to right in each of the panels of Fig.~3: 
particles that are less bound to the satellite are lost 
earlier. With time, dynamical friction gradually drags the satellites remnant closer to the centre of the host, towards the top of each panel, 
causing particles that were initially more bound to also be tidally stripped.

\begin{figure*}
\centering
\includegraphics[width=\textwidth]{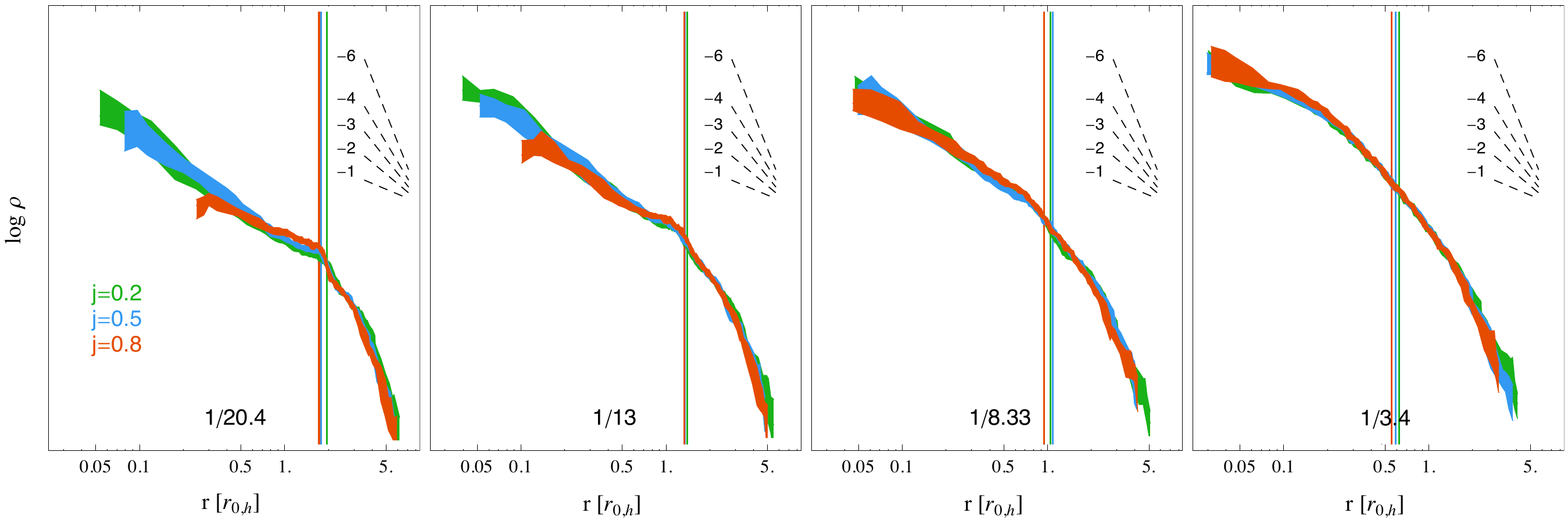}
\caption{The final density distribution of the deposited stars (the most bound $5\%$ of the satellite particles). Panels illustrate the effect of different virial mass ratios, as indicated in the bottom of each panel. Different colours in all panels are used for infall orbits with different initial circularity: $j\in\{0.2, 0.5, 0.8\}$, as in the legend in the leftmost panel. All profiles are normalised to the same total mass, showing that more massive satellites cause higher central densities, by depositing their most bound particles deeper in the host potential. Vertical coloured lines display the half-mass radius associated to each density profile.}
\label{density}
\end{figure*}

Grey points in the different panels of Fig.~3 illustrate the cases $j=0.5$ and 
${{r_{0,{\rm s}}/r_{0,{\rm h}}}={(r_{0,{\rm s}}/r_{0,{\rm h}})_{\Lambda {\rm CDM}}}}$ (runs A, B, C, E, F).
For each mass ratio, coloured lines display medians (with associated 1-sigma uncertainty) for the three runs with different initial circularities, 
as indicated in the leftmost panel.
At a given binding energy within the satellite $(E/\Phi_{\rm min})_{\rm sat}$, the distribution of energies within the host 
$(E/\Phi_{\rm min})_{\rm merg}$ is bimodal around the median. Particles with higher (lower) values were lost in the leading 
(trailing) condition \citep[e.g.,][]{KJ98,NA15}, when the remnant had approximately the corresponding median energy value. 
The distribution of particles in the scatter plots is also non-homogeneous, featuring
for example clear overdensities in the form of horizontal streaks. These represent collections of particles with 
very similar energies in the host, lost together in a coherent manner, which is what happens at pericentric passages.

For all mass ratios and for all
initial circularities, the least bound particles within the satellite are also the least bound particles within the merger product at the end of
the run. Their energies are distributed around the initial orbital energy of the satellite at infall.
Particles that were more tightly bound within the satellite are correspondingly more bound within the host,  with a clear gradient enforced 
by dynamical friction.  The existence of such a gradient is not new in the literature, and in fact an expression of the tendency of collisionless mergers 
to preserve the rank order of particles in energy \citep{SW78, SW80, JB88,PH09}. What is most interesting here, however, is that the slope of the same gradient is a strong function of the satellite mass. Only the most massive minor
mergers are capable of delivering their {stars} deep into the central regions of the host. In turn, the initial orbital 
circularity does not represent an important factor in this respect, with the differently coloured medians being practically indistinguishable in all panels. 


Figure~4 shows the spherically averaged density profiles of the stars deposited in the final merger product, with associated one-sigma uncertainty.  
Satellite masses grow from left to right and the colour coding in each panel is 
the same as in Fig.~3, indicating different initial orbital circularities. All density profiles are scaled so to integrate to the same total mass. 
First, it is evident that more massive satellites imply higher central densities, which is a direct outcome of the gradient observed in Fig.~3 in energy-space. Vertical coloured lines in Fig.~4 show the half-mass radius $r_{0.5}$ of each stellar density profile. As a consequence of the gradient in the effectiveness of dynamical friction, half mass radii become smaller for higher satellite masses { (Appendix A explore the dependence of the half-mass radius
with the chosen tagging fraction $f_{\rm mb}$).} Second, the initial circularity 
is only important at low satellite masses, where the density profiles resulting from satellites infalling on more circular orbit display central density holes.
In turn, for massive satellites, the density profiles resulting from accretion events with different circularities at infall are practically indistinguishable. 
The reason for this is
explored in Section~5, which concentrate on the kinematics of the deposited material.
Finally, there is a trend in the structure of the density profiles with satellite mass: low mass satellites result in more clearly broken density profiles, with 
a radius in which their logarithmic slope evolves sharply; massive subhaloes deposit stars in apparently smooth density profiles, 
with a gentle and progressive steepening of the logarithmic slope. 


\begin{figure*}
\centering
\includegraphics[width=\textwidth]{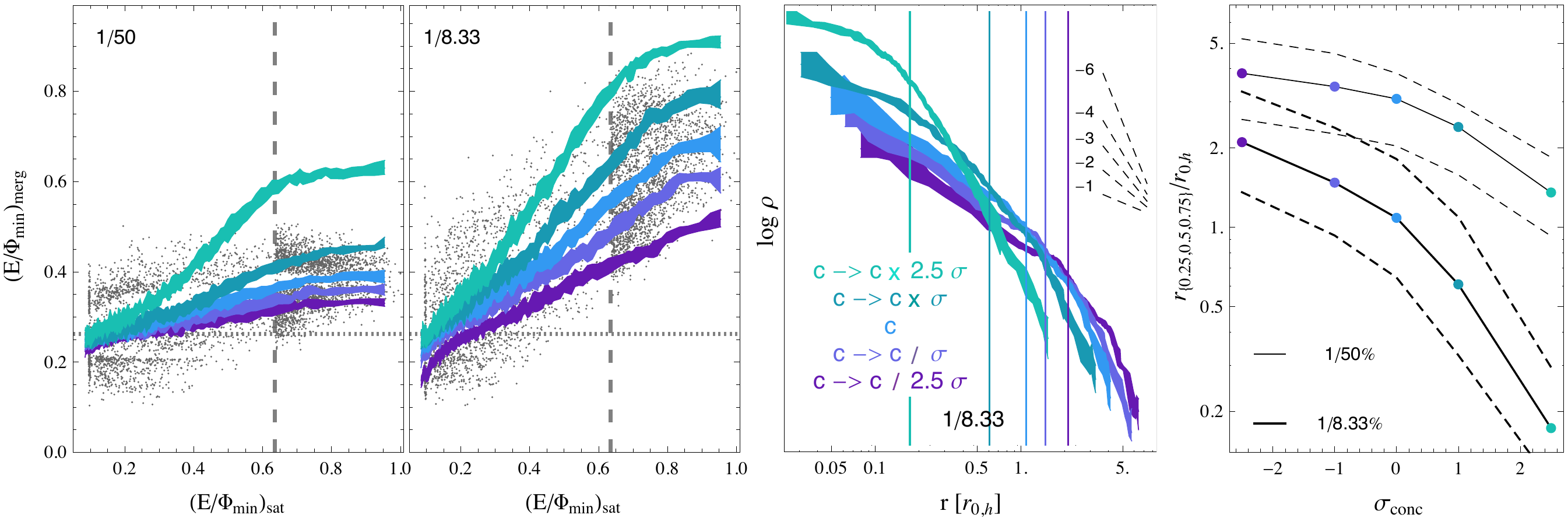}
\caption{The effect of satellite concentration. Different colours compare the final energy and density distributions of the stars deposited by satellites with different concentrations, probing an interval equivalent to $\approx \pm2.5 \sigma_c$ observed in cosmological simulations. Vertical lines in the third panel display the half-mass radii of each plotted density profile. {These are plotted as a function of the shift from the mean concentration in the rightmost panel, together with the radii $r_{0.25}$ and $r_{0.75}$, containing respectively 25\% and 75\% of the stars, for two values of the satellite-to-host virial mass ratio.} Satellites with higher concentrations sink further in before releasing their most bound particles, resulting in more concentrated, smoother accreted stellar profiles. As in Fig.~3, grey points refer to the case with average concentration and initial $j=0.5$, with an analogous change in sampling across the 5\% vertical line.}
\label{concentration}
\end{figure*}
\begin{figure*}
\centering
\includegraphics[width=\textwidth]{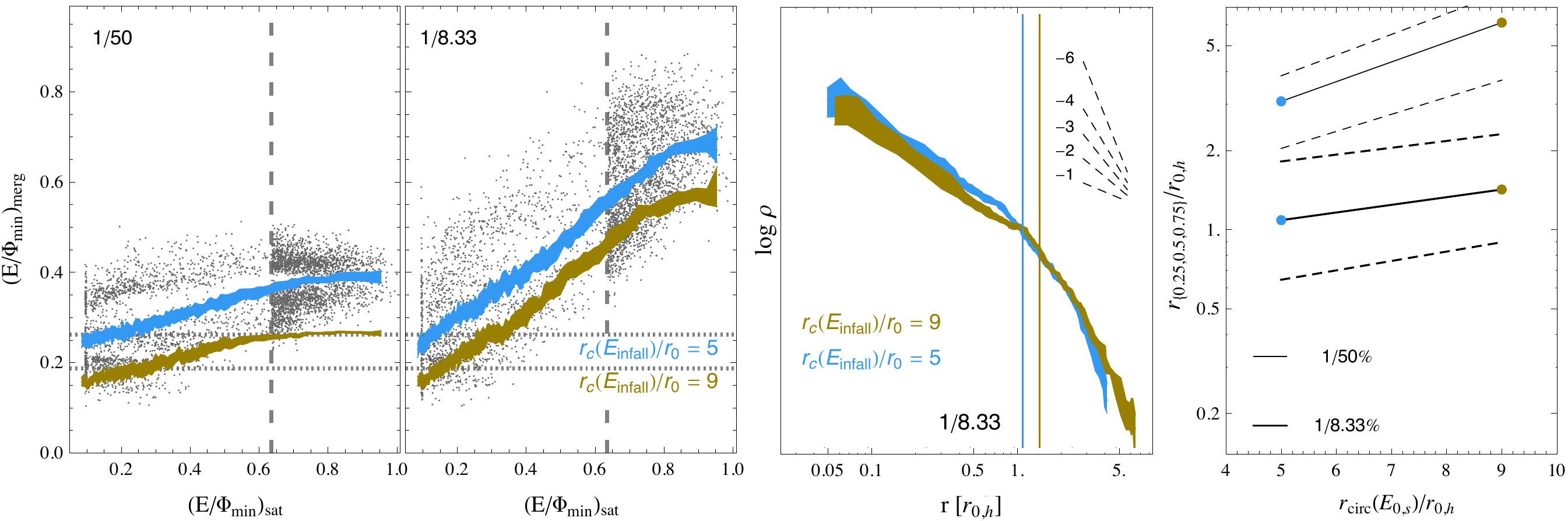}
\caption{The effect of the initial orbital energy at infall. Different colours compare the density and energy distributions resulting from satellites with different orbital energies, covering approximately the interval expected for recent accretions events on Milky Way like galaxies ($r_{\rm circ}/r_{0,{\rm h}}=9$) to accretions at intermediate redshifts $z_{\rm inf}\approx 2$ ($r_{\rm circ}/r_{0,{\rm h}}=5$). Vertical lines in the third panel display the half-mass radii of each plotted density profile. { These are plotted as a function of infall orbital energy in the rightmost panel, together with the radii $r_{0.25}$ and $r_{0.75}$, containing respectively 25\% and 75\% of the stars.} As in Fig.~3, grey points refer to the case with average concentration and initial $j=0.5$, with an analogous change in sampling across the 5\% vertical line.}
\label{orben}
\end{figure*}

\subsection{The effects of concentration and orbital energy at infall}

The results presented so far pertain to the median population of $\Lambda$CDM haloes, with runs A to F adopting the prescription 
${{r_{0,{\rm s}}/r_{0,{\rm h}}}={(r_{0,{\rm s}}/r_{0,{\rm h}})_{\Lambda {\rm CDM}}}}$. Figure~5 illustrates the effect of the scatter in the
concentration of the satellite haloes. As mentioned in Sect.~2, runs G and H adopt shifts to the halo concentration of the satellites corresponding 
to factors of approximately $\{1,2.5\}\times\sigma_{c}$, where $\sigma_{c}$ is the (logarithm of the) scatter observed in cosmological 
simulations \citep[e.g.,][]{AL14}.
The colour-coding is the same used in Fig.~1, so that the nominal run, with ${{r_{0,{\rm s}}/r_{0,{\rm h}}}={(r_{0,{\rm s}}/r_{0,{\rm h}})_{\Lambda {\rm CDM}}}}$ 
is always shown in turquoise, and is accompanied by the full particle scatter plot (grey points).

Satellites that are less concentrated than average (or that infall on hosts that are more concentrated than average) do not manage to
reach as deep into the host potential as their analogues with the same initial mass. Of course, this is a result of a quicker tidal stripping:
less concentrated satellites are less dense and lose mass at a quicker rate, which does not allow dynamical friction to consume a similar amount of orbital energy. The effect of concentration is quite marked on the resulting 
density profiles and half-mass radii of the deposited stars. The concentrated satellites contribute stellar populations with much higher
central densities, with smoothly falling density profiles, without sharp breaks. In turn, stars contributed by low-concentration satellites feature 
more marked breaks at large radii and their spatial distribution is considerably more diffuse.
As shown in Fig.~1, at fixed mass ratio, earlier accretions are characterised by higher values of the structural ratio $r_{0,{\rm s}}/r_{0,{\rm h}}$, 
i.e. lower values of the density contrast, and are therefore equivalent to mergers with slightly less concentrated satellites. 
From a purely structural point of view, this results in comparatively less deep contributions to the stellar halo, although 
this effect remains minor with respect to the physical evolution of the host, as it will be shown in the following Section.

Accretion events happening at higher redshift also differ for their average orbital energy, as 
$r_{\rm circ}/r_{0,{\rm h}}\approx c_{\rm h}$ \citep[e.g.,][]{LJ15}. As the concentration of the host increases monotonically with redshift, 
accretions at higher redshifts begin at slightly more bound initial normalised orbital energies. In order to investigate any effect of this 
evolution on the final location 
of the stellar debris, Fig.~7 compares runs with $r_{\rm circ}/r_{0,{\rm h}}=5$ and $r_{\rm circ}/r_{0,{\rm h}}=9$ (respectively, runs B E and I J). 
I find that a more loosely bound initial condition (i.e. $r_{\rm circ}/r_{0,{\rm h}}=9$, meaning a more recent accretion event) 
results in symmetrically more extended contributions to the stellar halo. Note in particular the apparent similarity of 
two colored median tracks, which are are essentially parallel to each other: a shift in the initial infall orbital energy results in a shift 
in the final mean energy of the deposited material.
Finally, there is perhaps a mild difference in the shape of the stellar density profile resulting from the two cases, with the higher-energy case 
$r_{\rm circ}/r_{0,{\rm h}}=9$ featuring a somewhat better defined density break.

\begin{figure*}
\centering
\includegraphics[width=.75\textwidth]{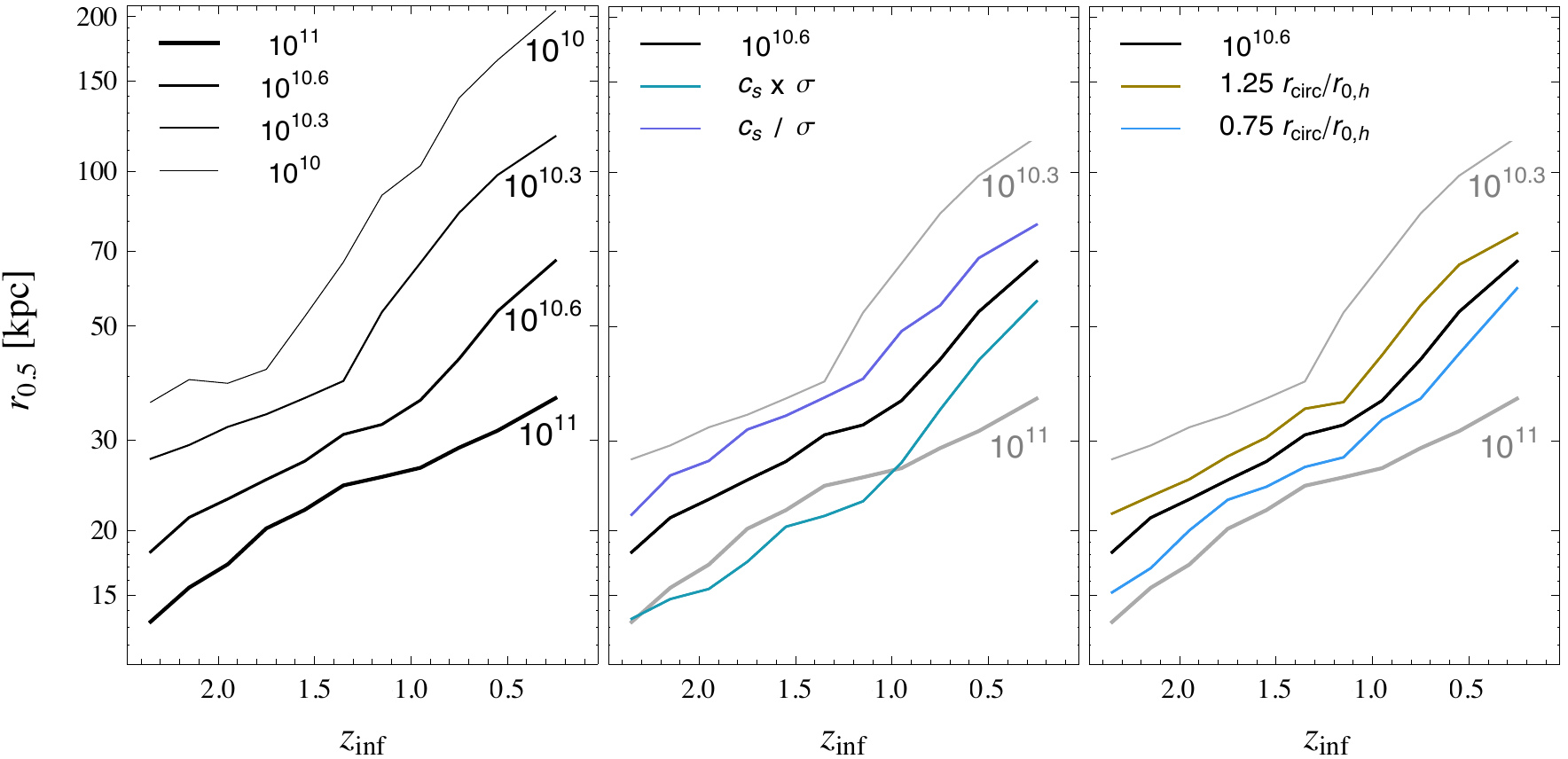}
\caption{The final half-mass radius $r_{0.5}$ for contributions to the stellar halo
of a Milky Way like galaxy, by satellites of different masses (lines of different thickness), infalling at different
redshifts $z_{\rm inf}$. Middle and right panels compare the pure influence of mass and redshift to the effects 
of satellite concentration and orbital energy at infall. }
\label{orben}
\end{figure*}

\section{Redshift evolution: Accretions onto a Milky Way like galaxy}

In this Section, I use the results described so far, laid out in structural dimensionless units, 
to derive scalings in physical units representative for accretions 
onto a Milky Way like galaxy. I have shown that the half-mass radius $r_{0.5}$ is practically 
insensitive to the initial orbital circularity of the satellite, and therefore it represents a useful 
quantity to explore the effect of the other ingredients at play. In particular, here I concentrate 
on satellite mass and infall redshift.

As mentioned in Sect.~2, structural parameters of average 
$\Lambda$CDM minor mergers at redshifts $z_{\rm inf}\lesssim 2.5$ lie within the area covered by 
the scatter in the concentration-mass relation at the present epoch. Therefore, I can derive the structural properties of accretion events 
at intermediate redshifts by interpolating the results of the runs presented here. As to physical scales, I adopt that, at $z=0$ a Milky 
Way like host has $M_{\rm vir,h}=10^{12}M_{\odot}$, and that at all times, it was in good agreement with the average properties 
observed in cosmological simulations. This assumption regards both its mass accretion history \citep[which I assume follows a median 
accretion history as compiled by][]{Fa10}, and its concentration \citep[which, for consistency, I assume follows the 
median relation of the Millennium haloes as compiled by][]{LG08}. Therefore, for example, the host halo currently 
has a characteristic radius of $r_{0,{\rm h}}(z=0)\approx21$kpc, while this was about $r_{0,{\rm h}}(z=2)\approx16.5$ kpc at higher redshift,
when its mass was $M_{\rm vir,h}\approx10^{11.55}M_{\odot}$. { I use the scale radius of the halo $r_0$ at the 
time of accretion to scale the dimensionless density profile of the deposited material. This implies I am ignoring the  
effect that any subsequent evolution of the host \citep[which has been shown to be limited, e.g.,][]{HB14} 
may have on the phase space coordinates of the deposited stars. }

Before describing results, it is fair to mention that these should be regarded only as educated estimates, 
due to the several working hypotheses on which they depend. In order of importance, I am assuming that: (i) the host
is `average' at all times since $z\approx 2.5$; (ii) the 
influence of any stellar component in the host on the dynamics of the minor merger can be neglected \citep[which may be worrisome 
because of the dynamical effects of the disk, see for example][]{JP04, AM15a}; (iii) satellites are strongly dark matter dominated, and the tagging criterion $f_{\rm mb}=5\%$
well describes their stellar components; (iv) once stars are deposited by each satellite, the host's mass growth has
a negligible effect on their density profiles, which also means that the host has not experienced major mergers since.

Under this set of assumptions, the left panel of Figure~7 shows the evolution of the half-mass radius $r_{0.5}$
with both the virial mass of the satellite (lines of different thickness in the left-most panel) and for different infall
redshifts. Note that, while in all previous plots I have used fixed values of the satellite to host virial mass ratio, 
here I am using the actual satellite's virial mass at infall. Therefore, as a result of the mass evolution of the host, 
different lines imply a mass ratio that increases slightly with redshift. 
I have already shown that more massive satellites deposit stars deeper into the host potential, and this is clearly visible in Fig.~7.
Additionally, Fig.~7 is showing a clear gradient with infall redshift $z_{\rm inf}$: for equal satellite masses, stars deposited by earlier 
accretions can now be found closer to the host's centre. This effect has also been seen
by \citet{AP14}, that measure that the stellar haloes of galaxies with a quicker assembly history (earlier halo formation time)
display steeper density profiles, and are therefore more compact. 

The driving effect of this evolution is in the size evolution of the host. 
The scale radius of the host's halo grows monotonically with redshift, implying that similar dimensionless radii are scaled to 
larger physical radii in the host for more recent accretions. Additionally, at higher redshift the host's concentration is also lower,
implying initially more bound energies at infall, the effect of which has been illustrated in Sect.~3.2. The combination of these two 
mechanisms is particularly important for the low mass satellites, where the counter-effect of dynamical friction has little influence.
Satellites infalling more recently are dragged more strongly by dynamical friction, as a result 
of their higher density contrast (see Fig.~1). However, this effect is practically negligible for those 
satellites with $M_{\rm vir, s}/M_{\rm vir, h}\lesssim 1/50$, where dynamical friction is almost negligible. 
Therefore, the half-mass radius of the stars deposited by the satellites with the lowest mass explored in Fig.~7
evolves by almost an order of magnitude between accretions at redshift $z_{\rm inf}\approx 2.5$ and very recent events.
{Note, however, that some of such low mass satellites with very recent accretion redshift might not have been 
entirely stripped of their stars at redshift $z=0$. This is an effect that is not explicitly considered in Fig.~7.}

After satellite mass and infall redshift, the leading factor in shaping stellar deposition is the satellite's concentration, the effect of which 
is shown by the middle panel of Fig.~7. A shift of a factor one-sigma in the satellite's concentration results in changes to the half mass radius 
of the deposited stars that is roughly similar to what a factor of $\approx 2.5$ in mass would cause. In turn, the average scatter in the 
orbital energy at infall \citep{LJ15} has a comparatively smaller importance, as shown by the rightmost panel.

\begin{figure*}
\centering
\includegraphics[width=\textwidth]{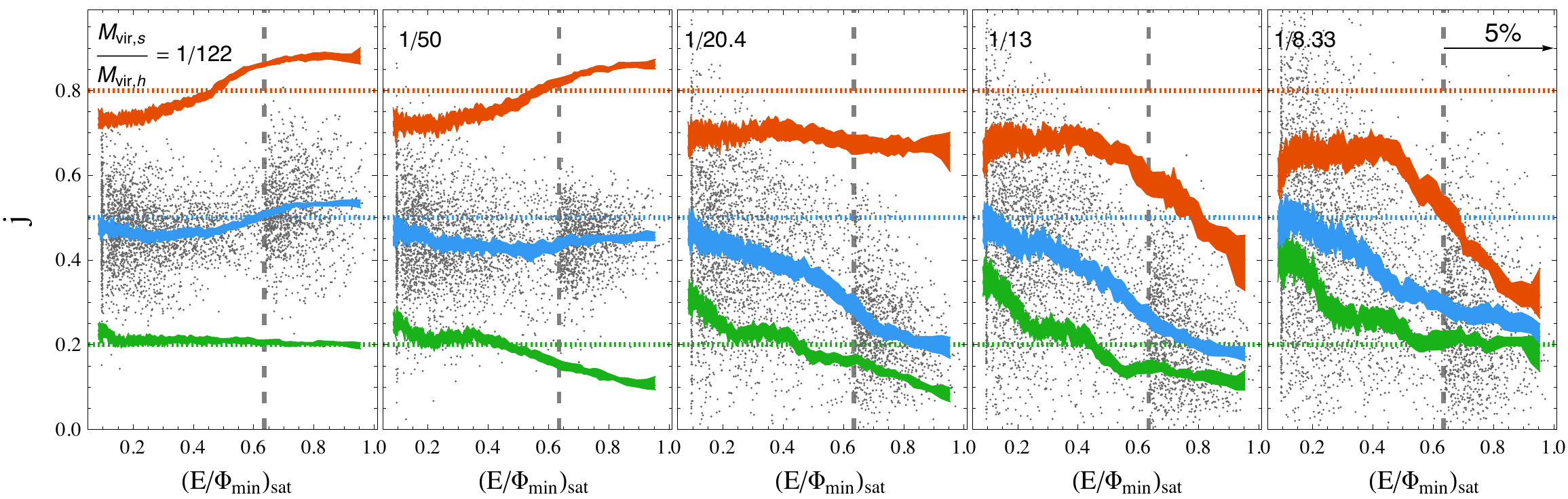}
\caption{Final circularities of the satellite's particles within the merger remnant. As in Fig.~3, the satellite's particles are ordered 
according to their binding energy within the satellite itself, prior to the merger.
The $y$ axis shows the orbital circularity of each particle within the host after it escapes the satellite, at the end of the simulation. 
Colour coding is analogous to Fig.~3, horizontal dotted lines show the 
values of the initial orbital circularity of the satellite at infall. Grey points refer to the case with $j=0.5$, with the same change in sampling (across the 5\% vertical line) adopted in Fig.~3 and~5.}
\label{jjfig}
\end{figure*}

\section{Kinematics of the contributed stars}

In this Section, I concentrate on the orbital properties and kinematical profiles of the material contributed by each accretion event.
This subject is closely connected to the orbital evolution of the satellite, as particles are lost with energies and angular momenta that
are close to the ones of the remnant at the time of their escape \citep[e.g.][]{KJ98,NA15}. Therefore, whether the orbits of the contributed stars 
are preferentially radial or circular mainly depends on the properties of the remnant's orbit during its disruption.

\subsection{Orbital evolution and radialisation} 

Unfortunately, it is not possible to assume that the orbital circularity of the remnant remains constant during the satellite's 
disruption for all initial mass ratios and circularities at infall. Processing by dynamical friction may cause significant evolution 
between infall and the time when stars are lost.
When used with a constant Coulomb logarithm, the dynamical friction formula derived by \citet{Cha43} suggests orbital circularisation.
However, historically, the comparison between the prediction of this simplified formula and results of actual N-body 
simulations has met with mixed success \citep[e.g.][]{VW99,JB00,TB01,Ha03}. Reasons for this are that: (i) in opposition to Chandrasekhar's 
framework, the Coulomb logarithm is an evolving quantity in a merger; (ii) material that has just been lost from the satellite 
can still contribute to the density wake, and therefore increase the rate of energy and angular momentum loss \citep[e.g.,][]{Fu06,Fe07}. 

Figure~8 explores the evolution of the orbital circularity of the remnant through the final circularity of its particles within the merger remnant.
The structure of Fig.~8 is analogous to the one of Fig.~3: the $x$ axis is the normalised binding energy of the satellite particles within the 
satellite itself, the $y$ axis shows the circularity of such particles in the merger remnant, at the end of the simulation. Stars are those particles
to the right of the vertical dashed line, which identifies the tagging threshold $f_{\rm mb}=5\%$. As in Fig.~3, coloured areas illustrate median values (with associated one-sigma uncertainty) for satellites infalling with different initial circularities, as shown by the horizontal dotted lines. 
As in Fig.~3, tidal stripping, dynamical friction and time all proceed from left to right in each panel. Therefore, median tracks with a 
positive gradient indicate a circularising progenitor, while a negative gradient identifies an evolution towards a more radial orbit.
As for Fig.~3, grey points within each panel illustrate the case  $j_{\rm inf}=0.5$.

There are a few points worth noticing. 
\begin{itemize}
\item{As it could be expected, the satellites particles that are least bound to the satellite itself 
cluster in general close to the initial orbital circularity of the progenitor at infall. However, with increasing satellite mass, 
the median circularity at low values of $(E/\Phi_{\rm min})_{\rm sat}$ departs from the nominal initial orbital 
circularity of the satellite. This is not an effect of dynamical friction, but the result of the
growing internal velocity dispersion of the satellite, which `dilutes' the imprint of the bulk orbital velocity. For example, 
the median circularity of the least bound particles in massive satellites infalling with $j_{\rm inf}=0.8$, grows lower than this figure.
Symmetrically, it gets higher than $j_{\rm inf}=0.2$ for those massive satellites infalling on almost radial
orbits.}
\item{In proceeding towards particles with higher and higher values of the binding energy $(E/\Phi_{\rm min})_{\rm sat}$, 
the scatter around the median circularity within the host decreases. This is a consequence of the decreasing bound mass of
the remnant, which implies a decreasing scatter in the kinematic properties of the particles at the time of shedding.}
\item{Low mass satellites that infall on quite circular orbits experience some mild circularisation. This happens
in both cases $M_{\rm vir, s}/M_{\rm vir, h}\in\{1/122, 1/50\}$ when infalling with $j_{\rm inf}=0.8$. Satellites of similar masses 
either do not experience significant evolution in their orbital circularity (like in the case $j_{\rm inf}=0.5$), or are dragged 
towards even more radial orbits (like for the case $j_{\rm inf}=0.2$ and $M_{\rm vir, s}/M_{\rm vir, h}=1/50$).}
\item{Satellites with higher masses, independently of their infall circularity, are uniformly dragged towards more radial orbits.
This is especially evident at intermediate satellite masses, $M_{\rm vir, s}/M_{\rm vir, h}\in\{1/20.4, 1/13\}$, where the scatter
due to the internal motions is not large enough to cover this median trend. This is in contradiction with the idea that dynamical friction generally 
circularises orbits. Under the assumptions of this work, I find that the most important contributors to the stellar halo are actually 
affected by dynamical friction in the opposite way. Recall, however, that it is not possible to exclude that a particularly 
massive stellar disk in the host may affect this conclusion.}
\item{For satellites that are massive enough, processing by dynamical friction can deprive the deposited stars of almost all memory of 
the initial orbital circularity of the satellite, as shown by the converging median tracks in the rightmost panel and explored quantitatively in the next Section. }
\end{itemize}
\begin{figure}
\centering
\includegraphics[width=.85\columnwidth]{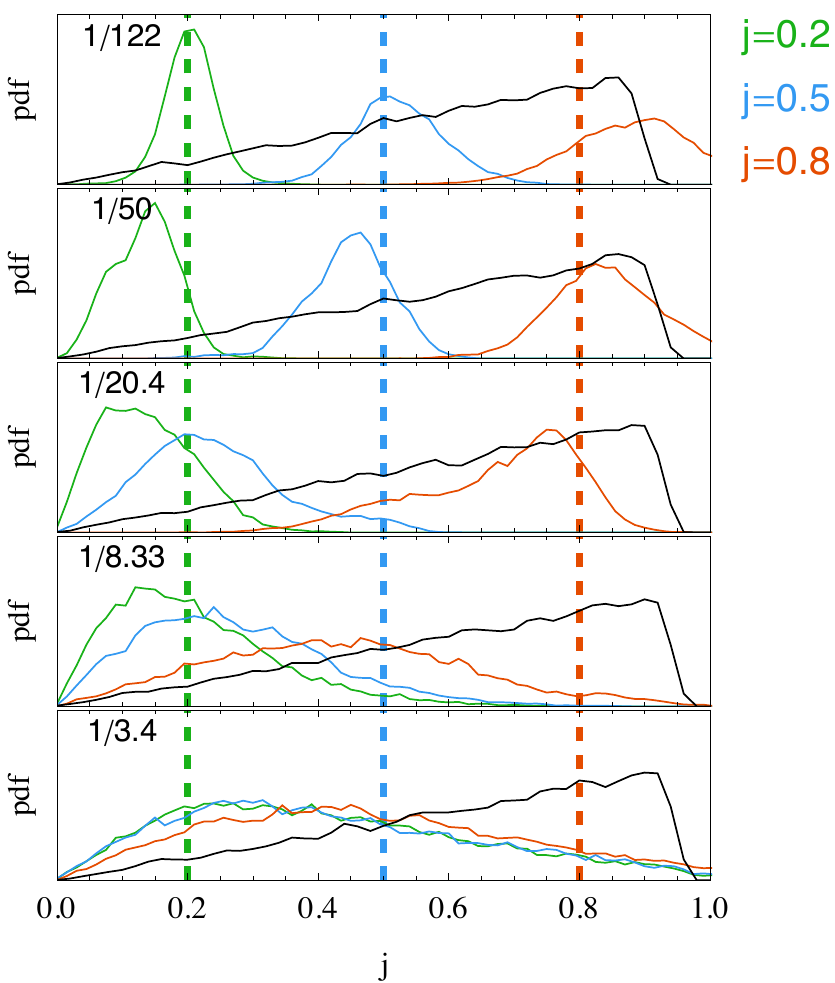}
\caption{Final circularity distribution of the most bound $5\%$ of the satellites particles. Dashed lines indicate the initial orbital circularity at infall. 
Black lines show the circularity distribution of an isotropic population ($\beta(r)=0$).}
\label{circularity}
\end{figure}

\subsection{Stellar circularites}

Figure~9 shows the implications of these results on the deposited stellar populations:
the different panels display the probability distribution of the final orbital circularities within the host 
for different satellite-to-host virial mass ratios. Vertical coloured lines mark the progenitors' orbital circularity at infall, for comparison.
As discussed earlier, the only satellites to deposit stars on orbits that are on average more circular than the infall orbit itself
(i.e. to experience orbital circularisation) are those with low-mass and with high circularity at infall. 
In these cases, the lack of stars with low angular momentum implies deposited stellar populations with central density holes, as it can be seen in Fig.~4. 
In all other cases, stars are either deposited with a mean circularity that is very similar to the orbital one at infall (in those cases in which the satellite
is not massive enough or the initial orbit is already very radial), or on orbits that are more radial that the one of the progenitor at infall. 
In particular, for massive satellites, their significant internal velocity dispersion and processing by dynamical friction 
make the circularity distribution of the deposited stars almost insensitive to the orbital circularity at infall. 
This is clear in the cases $M_{\rm vir, s}/M_{\rm vir, h}\in\{1/8.33, 1/3.4\}$, in which the different probability distributions
become similar.

Each panel of Fig.~9 also features a black profile: this shows the circularity distribution of a population
that (i) lives within the same gravitational potential, (ii) has the same energy
distribution as the deposited stars, (iii) has an isotropic dispersion tensor, $\beta(r)=0$ everywhere. Here,  
\begin{equation}
\beta(r)=1-{\sigma^2_t(r)\over2\sigma^2_r(r)}\ ,
\label{beta}
\end{equation}
is the anisotropy parameter, in which $\sigma_t$ and $\sigma_r$ are respectively the tangential and 
radial velocity dispersions. Note that the isotropic populations have a significant
fraction of members with high circularities. This is not the case for most stellar populations deposited 
by minor mergers. As a consequence, each single contribution to the stellar halo
represents a radially biased kinematical population, as it will be shown in Sect.~5.4. 
This is not automatically true for the full halo, that is for any superposition of a set of these contributions. 
Populations that retain some ordered rotation, if superposed with different alignment, will result in a less 
marked radial bias at the expense of bulk rotation. Whether the contributions from minor mergers retain 
ordered rotation, by preserving a fraction of the orbital angular momentum of their progenitor, is the subject of next Section.

\begin{figure*}
\centering
\includegraphics[width=\textwidth]{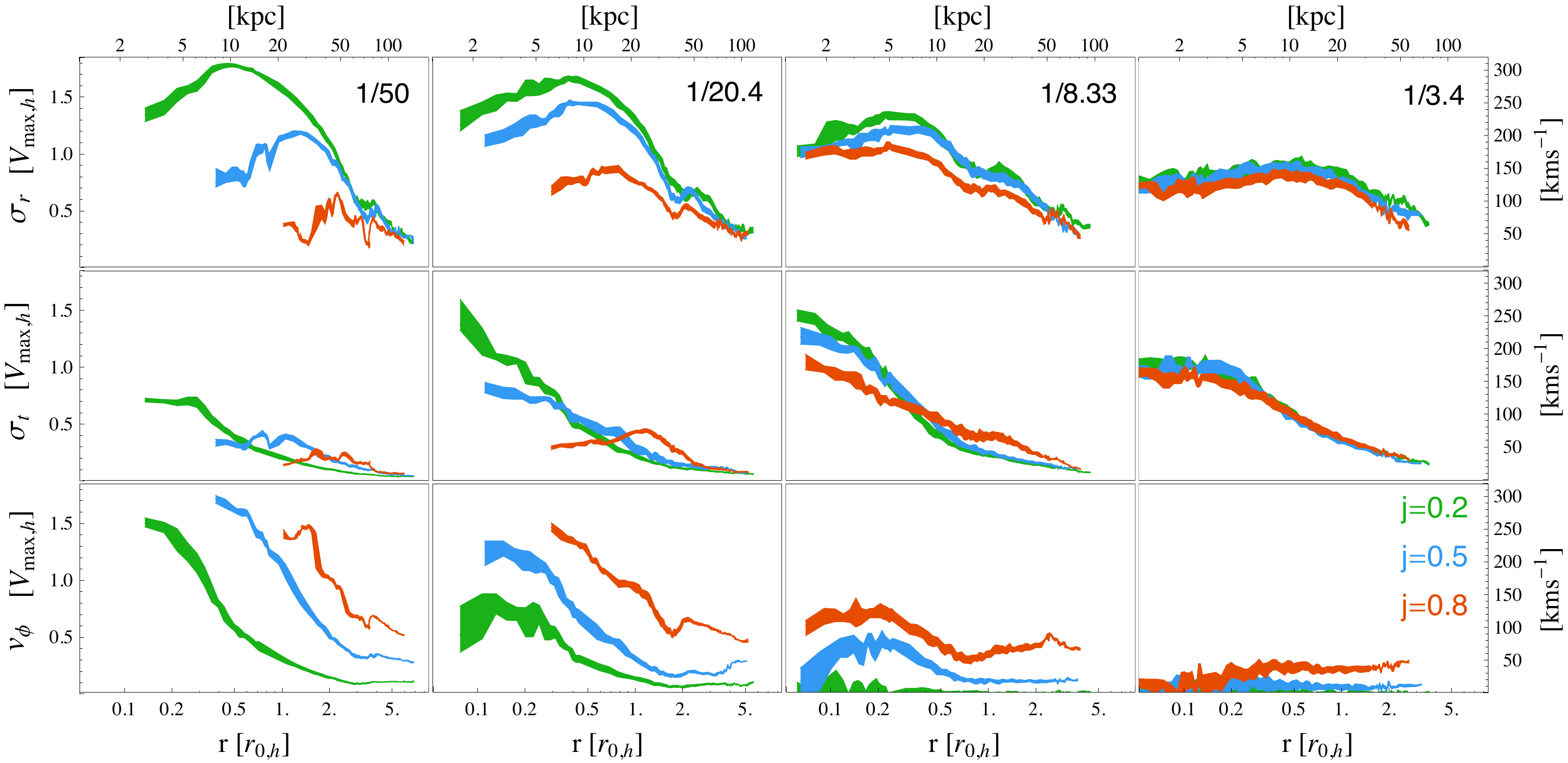}
\caption{An atlas of the kinematical profiles of the most bound $5\%$ of the satellites particles. The three rows of panels display respectively
radial velocity dispersion $\sigma_r$, tangential velocity dispersion $\sigma_t$ and ordered rotational velocity $v_{\varphi}$.}
\label{kinm}
\end{figure*}

\subsection{An atlas of kinematic profiles}

Figure~10 is an atlas of stellar kinematic profiles (runs B, C, E, F).
Dimensionless structural units, $\{r_{0,{\rm h}}, V_{\rm max,h}\}$, respectively the scale radius and maximum circular velocity
of the host halo, are shown in the bottom and left axes. Top and right axes are obtained by scaling the latter dimensionless quantities
for a median Milky Way like galaxy at redshift $z=0$, $\{r_{0,{\rm h}}, V_{\rm max,h}\}=\{21.1 {\rm kpc}, 173.1{\rm kms}^{-1}\}$.  
All profiles in Fig.~10 are related to concentric spherical shells. The top, middle and bottom row display respectively profiles 
for the radial velocity dispersion $\sigma_r$, the tangential velocity dispersion $\sigma_t$, and the ordered rotational velocity $v_\varphi$. 
This is the rotational velocity corresponding to the initial orbital motion of the satellite: $\varphi$ is the angular direction within the
orbital plane of the progenitor,  $\mathbf{\hat\varphi}= \mathbf{\hat{r}}\wedge\mathbf{\hat{J}_{\rm orb}}$, where 
$\mathbf{J_{\rm orb}}$ is the satellite's angular momentum at infall.

The clearest gradients are those displayed by the streaming velocity $v_\varphi$. At similar masses, 
satellites that infall on orbits with higher initial circularity deposit contributions
to the stellar with a stronger rotational support, as a result of the higher initial angular momentum. 
Rotational support decreases strongly with satellite mass, for two reasons: (i) a higher satellite mass implies 
a wider scatter in the escape condition of stars, resulting in a diminished kinematical coherence; (ii) orbital angular 
momentum is more efficiently lost through dynamical friction.

Both streaming velocity $v_\varphi$ and the two components of the velocity dispersion, $\sigma_r$ and $\sigma_t$,
become more and more insensitive to the satellite's orbital circularity at infall with increasing mass.
This is a direct consequence of the behaviour shown by Figs.~8 and~9:
processing by dynamical friction becomes more and more efficient with satellite mass, so that, together with the density profiles,
the final kinematical profiles of the stars deposited by the most massive minor mergers bear hardly any memory of the initial conditions
of the accretion event. 

Other gradients but can be understood in terms of two basic ingredients:
(i) random motions of the deposited stars around the ordered motion imprinted by the remnant's orbit increases with mass,
(ii) the orbital energy within the host of stars deposited by satellites of higher masses is lower.
For example, let us concentrate on the evolution with mass of the radial and tangential velocity dispersion profiles.
\begin{itemize}
\item{A strongly radially biased population is characterised by a high radial velocity dispersion. This directly results 
from the mean orbital motion connected with such radial orbits, rather than from the scatter in the orbital properties
of each stars around the mean. As a consequence, the radial velocity dispersion of the stellar populations contributed by the 
most radial accretion events decreases with satellite mass. In particular, Fig.~10
implies that, in absence of other information, stars that move at high velocity on nearly radial orbits 
are preferentially deposited recently by low mass satellites.}
\item{In turn, the radial velocity dispersion of the material deposited by satellites infalling on orbits with high-circularity 
is mainly a result of the scatter in their specific orbital properties. Therefore, this should increase with satellite mass. 
However, increasing the satellite mass causes a reduction of the mean orbital energy, which reverses the trend between 
$M_{\rm vir, s}/M_{\rm vir, h}=1/8.33$
and $M_{\rm vir, s}/M_{\rm vir, h}=1/3.4$, where $\sigma_r$ slightly decreases. }
\item{A similar reasoning applies to $\sigma_t$ and to its evolution with mass. Increase of random motions at the expense of
ordered rotation is responsible its growth with mass while at low satellite masses. Decrease in the mean energy justifies 
the opposite trend at the high-mass end.}
\end{itemize}
%

\subsection{The influence of satellite concentration}

Figure~11 shows how the scatter in the mass-concentration relation affects the kinematic profiles of the 
accreted stars. As mentioned earlier, the most concentrated satellites (or the satellites that infall on the less 
concentrated hosts) survive longer because of the highest 
density contrast, and require dynamical friction to drag them closer to the host centre before they can be 
efficiently stripped of their most bound particles (see Fig.~5). As a direct consequence, the average
orbital energy of the deposited tagged material decreases with increasing satellite concentration, 
and a larger fraction of the angular momentum is lost. 

This is clearly seen in the profiles displayed in Fig.~11. (i) The ordered rotation $v_\varphi$ retained within the stellar populations
decreases monotonically with the satellite concentration. 
(ii) Radial and tangential velocity dispersion profiles are clearly ordered with concentration. Both these points are especially true for 
the material deposited closer and closer to the host's centre, which has experienced an increasing amount of dynamical friction
and has more and more bound energy values within the host.

Despite the gradual loss of ordered motion with growing satellite concentration, 
the radial bias in the different cases remains very similar (bottom-right panel). While the evolution of both $\sigma_r$
and $\sigma_t$ is significant, this takes place in a coherent manner, so that their ratio is largely insensitive
to concentration, and each single contribution remains largely radially biased. As already mentioned earlier, it should be kept in mind 
that a superposition of populations rotating in different directions may be less radial than the single components. 
This effect, though, will not be significant when the major contributions come from satellites that have had 
their orbital angular momentum largely consumed by dynamical friction.

\begin{figure}
\centering
\includegraphics[width=\columnwidth]{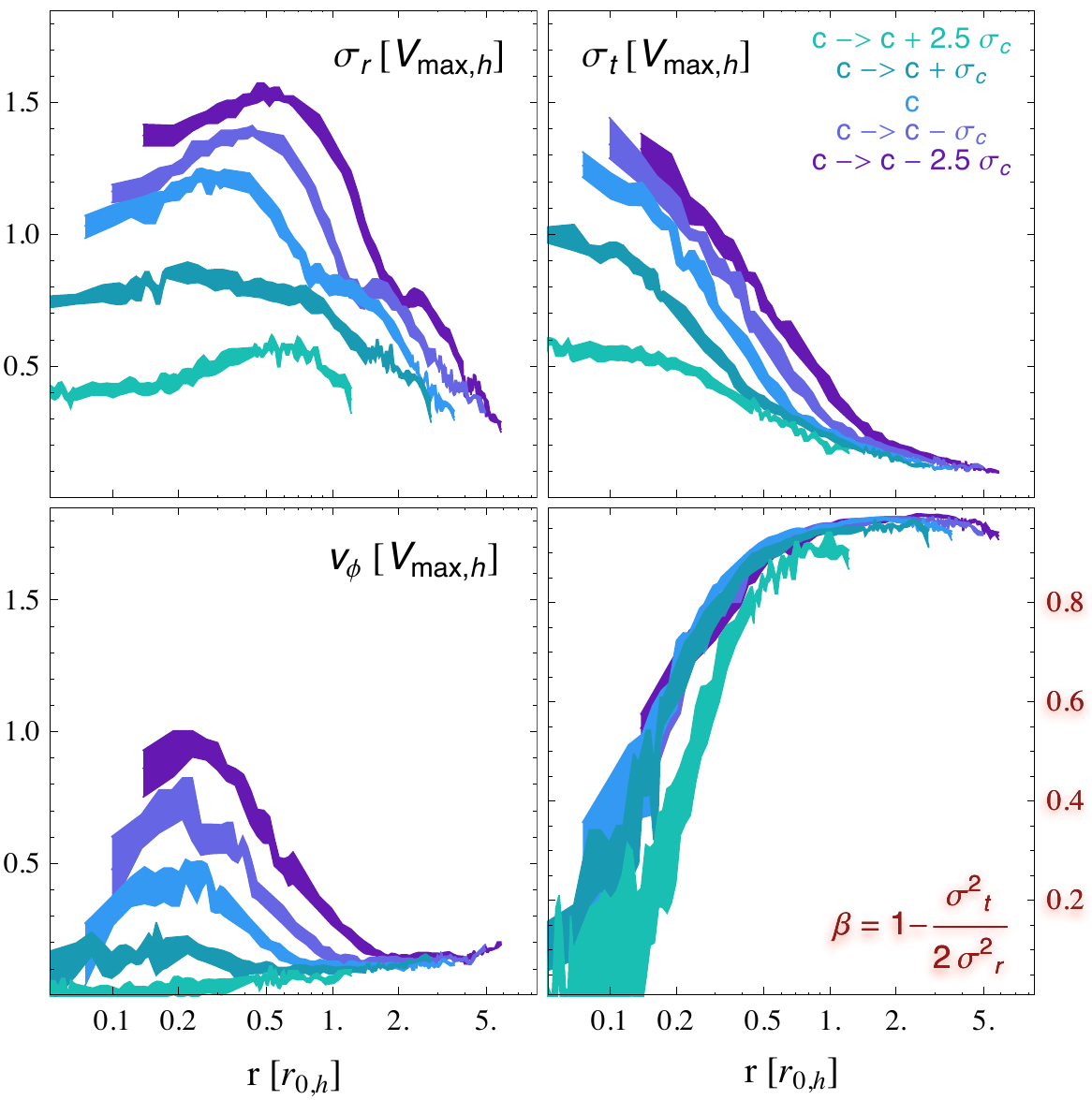}
\caption{The effect of the satellite's concentration on the kinematics of the deposited material, assuming $f_{\rm mb}$ selects the most bound $5\%$. 
Displayed profiles pertain to runs E and H, with  $M_{\rm vir, s}/M_{\rm vir, h}=1/8.33$ and $j_{\rm inf}=0.5$.}
\label{kinmc}
\end{figure}

\section{Discussion and Conclusions}

Without aiming to produce realistic examples of galactic stellar haloes, here I have investigated the influence of
the different physical ingredients that shape the contributions of each single accretion event.
This exercise is a useful first step towards the interpretation of results of both recent numerical investigations into the 
detailed properties of stellar haloes \citep[e.g, ][]{MA06, AC10, AC13, Ti14, AP14, RG15}; and observational
campaigns targeting the outer faint envelopes of our own and nearby galaxies 
\citep[e.g., ][]{Ju08, Se11, AD11, AD14, KG12, RI14, vD14, DP15, AM15b, IT15}.

{ \subsection{Limits of this study}

There are several limitations to the present study.  
\begin{itemize}
\item{
Some derive from the simplifying assumptions of ideal, spherical
hosts and satellites. The most evident in this category is the lack of a disk 
component in the gravitational potential of the host, which is known to have a role in the general process of tidal disruption. First, it accelerates
the disruption of subhaloes through disk shocking \citep[e.g.,][]{ED10}, and second, it contributes an additional
torque that tends to align satellites with the disk plane \citep[e.g.,][]{JP04}. Ignoring these two effects, however, does
not invalidate the analyses of this work for the following reasons. First, the contribution of tidal shocking is only crucial to the evolution 
of subhaloes with low virial mass, in a regime that is not relevant here \citep[$M_{\rm vir, s}\lesssim 10^9 M_\odot$, ][]{ED10}, and 
is of little importance for those subhaloes that actually contribute the largest amount of ex-situ material to the stellar halo of Milky Way like galaxies.  
Second, I have not addressed `local' properties of the accreted halo, or tried to explore the differences of properties 
along particular directions. The presence of the host disk may induce differences in the accreted halo along directions oriented 
differently with respect to it \citep[e.g.,][]{AM15a}, and it is clear that this paper cannot address this. In fact, I have concentrated 
on averages obtained in spherical shells, which the presence or absence of a stellar disk leaves largely unaffected.}

\item{Other limitations follow more directly from the use of a particle tagging {technique. These} are shared 
with those studies that have addressed the build up of the accreted stellar halo within a similar framework \citep[e.g.,][]{BJ05,AC10}. 
The major implicit assumptions that lie within the tagging technique itself are that (i) the gravitational influence 
of the stellar component of the satellite and (ii) the detailed morphological properties of the satellite can both be ignored. 
However, it is known that structural properties can significantly affect the the post-infall evolution of the satellite \citep[e.g.,][]{ED10b, SK11, Ch13}, 
and it is therefore worth checking in which regime a particle tagging technique can be safely adopted.

The assumption (i) is clearly broken in the case of a satellite featuring a baryon-dominated, concentrated central component, 
like a central bulge. Density in such region would be significantly higher than in embedding NFW halo, and therefore such 
component would be more resilient to tidal stripping. 
Prolonged dynamical friction might affect the deposition of the stellar material, possibly introducing differences from 
what predicted by a particle tagging implementation. For example, it follows that particle tagging is not directly viable 
{to study the accreted stellar haloes of massive galaxies. These receive substantial contributions by satellites with virial masses comparable to the one of the Milky Way, 
characterised by baryon-dominated centers which are more dense than their NFW haloes.}
In turn, the stellar halo of a Milky Way like galaxy is contributed by satellites with 
$M_{\rm vir, s}\lesssim 10^{11} M_\odot$, which have been shown to be dark matter dominated {at all radii} \citep[e.g.,][]{MW09,NA11,MC14,JV15}. 

The assumption (ii) is broken when the actual orbital structure of the satellite stars is significantly different from the approximately 
isotropic, pressure supported central regions of an NFW halo. This is the case when the satellite features a rotationally supported 
thin-disk satellite, which is known to respond to tides differently from a spheroid \citep[e.g.,][]{ED10, Ch13,JP10}. {In such a case,
the predictions of a particle tagging technique would not be reliable}. For example, \citet{ED10b} points out the differences 
in the post infall evolution of a disky satellite that spins with a prograde/retrograde alignment and \citet{Ch13} shows that
the removal of the satellite's stellar component is substantially more efficient when the morphology is the one of a thin stellar disk
rather than a spheroid. However, the dwarf members of the Local Group display very limited rotational support, 
with quite low values of the dimensionless parameter $v/\sigma$ \citep[e.g.,][]{LM01,AM12,CW15}. Even in those cases in which a velocity 
gradient is detected, their morphology remains far from the one of thin disks, and is substantially puffy \citep[see e.g.,][]{vdM01,RL12}.
\citet{AC10} has shown explicitly that both density and kinematics of the dwarf Spheroidal satellites of the Milky Way are well described 
by the most bound particles of an NFW halo.}

While the arguments above support the robustness of the results of this work, the present objective remains to identify and
explain the main correlations and physical mechanisms, rather than to provide precise predictions for any individual halo. 
Here I have concentrated mainly on the global properties related to the process of mass deposition, like the mean energy or 
the half-mass radius, and on the kinematics of the different contributions. 

\end{itemize}}

\subsection{Summary}

{Each individual contribution to the accreted stellar halo is defined within a parameter space of high dimensionality.
For Milky Way like masses and below only a handful of major contributions shape the resulting accreted halo, a number that is not high enough to allow for 
the properties of such haloes to converge. In this regime, therefore, the halo formation process remains highly stochastic, resulting in significant halo-to-halo scatter. 
Here, I have adopted a simplified approach to explore and classify the main degrees of freedom of the problem, and to deconstruct their influence in 
shaping stellar deposition by individual satellites. }Namely, I have concentrated on the
mass ratio of each accretion event, the internal structure of the satellite itself (in the form of its concentration and of the `size' of 
its stellar component, $f_{\rm mb}$), the infall redshift and the properties of the orbit at infall (its energy and circularity).

Main results are summarised below.
\begin{itemize}
\item{Massive satellites sink deeper into the gravitational potential of the host before stars are lost, so that these are contributed 
at smaller radii within the host. In turn, low-mass satellites survive much longer, but dynamical friction is not capable 
of dragging them within the innermost regions, with a clear segregation \citep[see also][who find clear 
evidence for this segregation in the accreted stellar haloes of the Illustris galaxies]{RG15}. Depending on their initial orbital circularity at infall and their concentration, satellites with 
$M_{\rm vir, s}/M_{\rm vir, h}\lesssim1/50$ may still display bound remnants at the end of the simulations (equivalent to 15 Gyr for a host
with $M_{\rm vir, h}=10^{12} M_{\odot}$).}
\item{Satellites that are accreted at higher redshift deposit their material at smaller radii within the host, as a result of the fact that 
the host was physically smaller at that time and that the orbital energy at accretion was lower.}
\item{The mass-concentration relation has a scatter that is wide enough to invert the ordering with mass: satellites that are one-sigma
more (less) concentrated than average can deposit their stars at radii that are closer in (further out), as a factor of $\approx2.5$ 
in mass (and average concentration) would imply. In turn, the scatter in the distribution of orbital energies at infall is not as important.}
\item{Dynamical friction can imply a marked evolution in the orbital properties of the remnant, shaping both density distribution and kinematics
of the deposited material.}
\item{Low-mass satellites infalling on high-circularity orbits experience some mild circularisation, and result in density profiles that feature 
central density holes. The dominating effect of dynamical friction on satellites with $M_{\rm vir, s}/M_{\rm vir, h}\gtrsim1/20$ is to 
quickly radialise their orbits, up to erasing memory of the initial infall circularity.}
\item{Material deposited by low mass satellites retains a significant amount of ordered rotation and, because of the extended orbits,
also features high radial velocity dispersion. In turn, angular momentum is consumed and diluted in the more massive accretion events,
resulting in almost non rotating contributions, with a strong radial bias.}
\end{itemize}

Finally,  {minor mergers have been suggested as a driver of the size evolution of}
massive elliptical galaxies with redshift {\citep[e.g.][and references therein]{RB09,TN09,RF10,vD10,Os10,Os12}}.
Indeed, here I am showing that mergers with {increasingly low} virial mass ratios result in 
stellar deposition  {increasingly large} radii around the host. This is a result of the switching 
balance between the rate at which satellites are stripped by tides and the rate at which they
sink trough dynamical friction. For hosts and satellites that populate the regions of parameter
space that is interesting for a $\Lambda$CDM universe, dynamical friction is faster than stripping 
for the massive minor mergers, resulting in stellar deposition within the inner regions of the host.
On the other hand, and despite the slow rate of tidal stripping, dynamical friction is too slow 
to drag the low mass satellites towards the centre of the host. Similarly, orbital evolution of the 
remnant is only effective at the high-mass end, where it can operate significantly before stars are 
lost. {In these cases}, it radialises the orbit of the progenitor satellite, with consequences on the 
kinematics of the deposited population, but also with possibly interesting effects in shaping 
the morphologies of low surface brightness tidal features \citep[][]{NA15,HJ15}.

\section*{Acknowledgments}
It is a pleasure to thank Annalisa Pillepich and Glenn van de Ven for stimulating conversations, 
Simon White for insightful comments and the anonimous referee for a constructive report. 
The Dark Cosmology Centre is funded by the DNRF.

\appendix

\section{Tagging fraction and half-mass radii}
\begin{figure}
\centering
\includegraphics[width=.7\columnwidth]{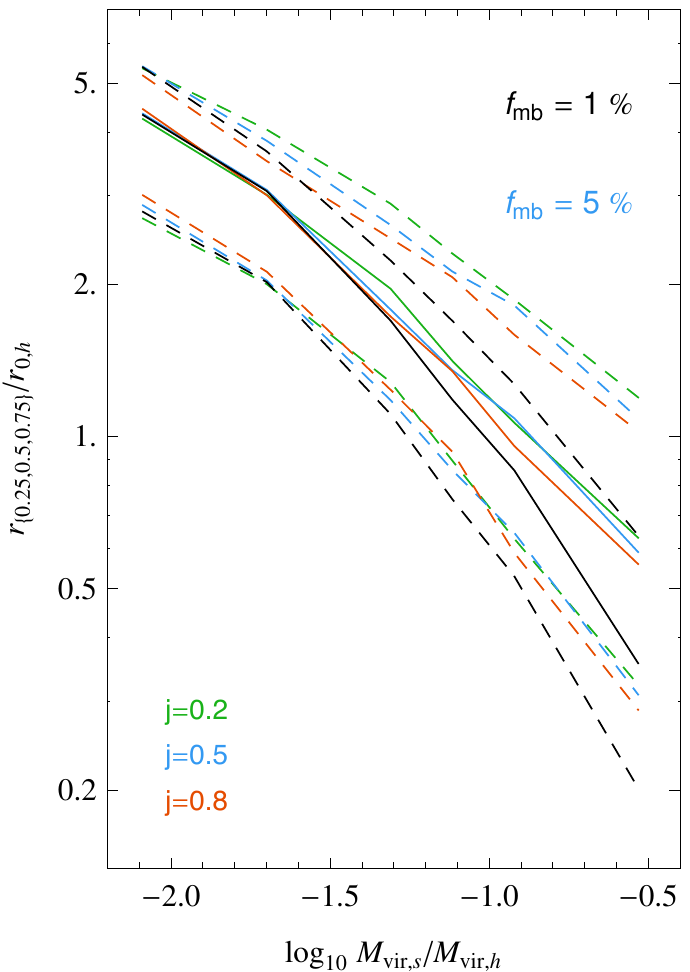}
\caption{The 25\%-, 50\%- and 75\%- mass radii for the density profiles generated assuming $f_{\rm mb}=5\%$ (profiles coloured according to circularity at infall, like in Fig.~3 and~4) and $f_{\rm mb}=1\%$ (black profiles, $j=0.5$). }
\label{circularity}
\end{figure}

{
Fig.~A1 displays the dependence on the virial mass ratio of the minor merger, $M_{\rm vir, s}/M_{\rm vir, h}$, of the radii 
containing $\{25\%, 50\%, 75\%\}$ of the contributed stellar mass, respectively $\{r_{0.25}, r_{0.5}, r_{0.75}\}$.
Additionally, Fig.~A1 shows the effect of lowering the tagging fraction, from the nominal value 
used throughout the paper, $f_{\rm mb}=5\%$, to $f_{\rm mb}=1\%$ (black profiles, for an initial circularity of $j=0.5$). 
For the satellites with the lowest mass, this does not introduce any noticeable difference. This follows from the fact that 
dynamical friction is not capable of consuming any more energy within the interval of time that separates the loss of the 
5\% particles and the escape of the 1\% particles.
However, more massive satellites do experience additional dynamical friction within this interval of time, and therefore 
density profiles generated using $f_{\rm mb}=1\%$ have smaller characteristic radii.
}


\end{document}